# Knowledge-based Expressive Technologies within Cloud Computing Environments

Sergey V. Kovalchuk, Pavel A. Smirnov, Konstantin V. Knyazkov,
Alexander S. Zagarskikh, Alexander V. Boukhanovsky[1]

**Abstract.** Presented paper describes the development of comprehensive approach for knowledge processing within e-Sceince tasks. Considering the task solving within a simulation-driven approach a set of knowledge-based procedures for task definition and composite application processing can be identified. This procedures could be supported by the use of domain-specific knowledge being formalized and used for automation purpose. Within this work the developed conceptual and technological knowledge-based toolbox for complex multidisciplinary task solving support is proposed. Using CLAVIRE cloud computing environment as a core platform a set of interconnected expressive technologies were developed.

## 1 Introduction

Today a lot of complex e-Sceince [1] tasks are solved using computer simulation which usually requires significant computational resources usage. Moreover the solutions, developed for such tasks are often characterized by structural complexity which causes a lot of different resources (informational, software or hardware) to be integrated within a single solution. The complexity of the solutions grows as the multidisciplinary tasks are considered. Today's common approach for building composite solutions is based on Service-Oriented Architecture [2] which forms the basis from interconnection of services and hiding their complexity behind their interfaces. Interconnection of the services within complex tasks is usually implement-ed in a form of workflow (WF) structures [3] which exploits graph-based structures to describe interconnection of used services. On the other hand, today the Cloud Computing [4] concept is developed as a business framework for pro-viding on-demand services supporting resources' consolidation, abstraction, access automation and utility within a market environment. To support the consolidation and abstraction description of available resources should be provided; automatic services composition requires the tool for composite application management; utility and market properties should be supported with semantic domain-specific description. Thus cloud computing platforms should provide the domain-specific user-oriented tools for expression of descriptive artifacts.

[1] S.V. Kovalchuk (✉), P.A. Smirnov, K.V. Knyazkov, A.S. Zagarskikh, A.V. Boukhanovsky
Saint-Petersburg National University of IT, Mechanics and Optics, Saint-Petersburg, Russia
e-mail: kovalchuk@mail.ifmo.ru

Finally, today Problem Solving Environment (PSE) [5] brings a set of domain-specific tools together to solve the proposed domain problems. Still this approach requires a set of knowledge from different domains being available for the user of a PSE. Intelligent PSE (iPSE) [6] tries to extend the PSE concept with knowledge-based using formalized knowledge within these domains.

In the presented work the conceptual and technological solution for comprehensive knowledge-based support of human-computer interaction within the process complex e-Science task solving is described. The key goal of the work is to organize different knowledge-based technologies within a continuous solution which support the integration process within cloud computing environment and enable automatic solving of technological issues.

## 2 Knowledge-based e-Science Technologies

The process of solving e-Science tasks is strongly related to the knowledge processing. This point of view (see fig. 1) can discover a set of features:

1. Today the development of global networking technologies within the Internet makes the international scientific *society* an important source of knowledge. This trend causes to appear a concept of Science 2.0 [7] which gives the important role to the global collaboration of scientists. As a result all the knowledge which is used to define the e-Science tasks can be considered as the knowledge of society. Moreover all the results obtained during the task solving process also could belong to the society and extend its knowledge.
2. The implicit set of knowledge can be *obtained* for the further utilization using different approaches. The most important of them are: explicit knowledge formalization performed by or within collaboration with experts and automatic analysis of tests, data and experiments published by the society.
3. *Formalized knowledge* within e-Science tasks can be divided into three main groups: a) domain-specific knowledge, which describes the specificity of the task within the particular problem domain(s); b) IT-related knowledge which is used for the automation services searching, tuning and calling; c) system-level knowledge which supports the organization of simulation and data analysis. Today the idea of system-level science [8] appears which is focused more on comprehensive exploration of the system rather than on running particular pro-cedures. Thus all these knowledge should be interconnected and coordinated around the process of simulation-based exploration of particular system.
4. The formalized knowledge can be used to *automatic support* of different procedures during the problem solving process. This knowledge allows designing human-computer interaction process in a way most suitable for the user, allowing to obtain all the necessary information to build, execute the composite application and analyze its results. Considering a WF-based technology there should be a) a set of tools for building and monitoring the WF using domain-specific concepts; b) a set of technologies which simplify the interpretation and execution of the WF by the use of knowledge on service usage.
5. As a result there should be a set of technological tools (domain-specific languages (DSL), editors, knowledge bases) which allows different users of the system (end-users, experts, developers, IT specialists etc.) to *express their knowledge* in a form which a) can be easily used by the user; b) will be understandable to any other user with the same knowledge background; c) can be interpreted automatically by the software which performs computing.

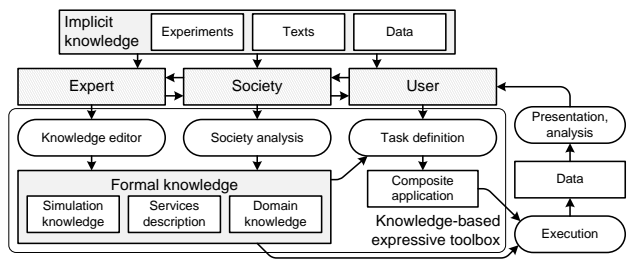

**Fig. 1** Knowledge-based view to e-Science tasks

The proposed knowledge-based expressive toolbox can be defined as a set of technologies which supports the process of composite application building, execution and results' analysis. The knowledge base which incorporates the knowledge of different kind is a core of this technological toolbox. The toolbox includes the following classes of technology available for the users of different classes: a) knowledge bases as a core integration technology; b) a set of DSLs; c) graphical user interfaces; d) performance models describing the execution of services.

## 3 Expressive Technologies

Presented approach is developed as an extension to the iPSE [6] concept which provides knowledge-based support of the problem solving process. Considering main operation which requires knowledge expression and interpretation the following procedures and modules can be defined within the computational environment built within a framework of this concept (fig. 2a). Each of these procedures can be associated with a specific descriptive language (with textual or graphical notation) which can provide a) support of the expression process by the use of knowledge-based tools; b) interpretability for the purpose of underlying simulation automation. These languages (fig. 2b) can be interconnected within a continuous stack with relation a) to the domain-specific entities within knowledge base; b) objects on different abstraction levels of the complex e-Science solutions. The expressive technology for each level includes a pair of tools: a) for expression of knowledge on corresponding level; b) for interpretation of these expressions.

1. Services (including computing services, data services, interfaces to the specific devices, observation sources etc.), available within computational environment might be described with a set of knowledge which defines the interface of service, its domain-specific interpretation and usage of support procedures. This description can be implemented using declarative DSLs. As the main WF representation is usually implemented using abstract service description, the mapping of abstract workflow (AWF) onto particular services should be done during the WF interpretation. Also the description of the services might support the execution monitoring. Still the most significant meaning of this part of expressive toolbox is providing domain-specific information on the user level. The descriptive language on this level should include structures for service and data structures semantic description.
2. Composition of the services is usually implemented in a form of AWFs, where particular services call is specified by the description of service's type and input/output flows. The interpretation of the WF includes mapping to the particular services, calling of these services and execution monitoring. The WF development requires set of new technologies to be developed. E.g. WF management systems (WFMS) provide different tools for service discovery and usage. Also the language for composition of abstract services should be developed.
3. The process of system's exploration is usually focused on the properties of specific domain objects which can be explored through the simulated model. The description of such object can be translated into AWF form. Nevertheless the semantic structure of simulated object can be considered as a separate entity which can be defined by the domain expert and used for further model-based exploration and hiding the complexity of the underlying WF. Thus, this structure (predefined simulation WF and its semantic interpretation) could be considered as a part of domain-specific knowledge. This level presents interconnection between domain-specific and IT knowledge.
4. The description of domain-specific object is a core entity for the simulation-based system exploration. The composition of such object can be interpreted a) as a semantic description of the system which is explored within the current task; b) as a basic structure for complex simulation WF composition which can be further translated through the levels 3 to 1. Moreover the semantic description of the system can be used in conjunction with the spatial/temporal basis of the simulation. This type of integration allows switching from the procedural point of view (where procedures are equivalent to the service calls) to the simulation-based point of view with automatic construction of the procedural WFs.

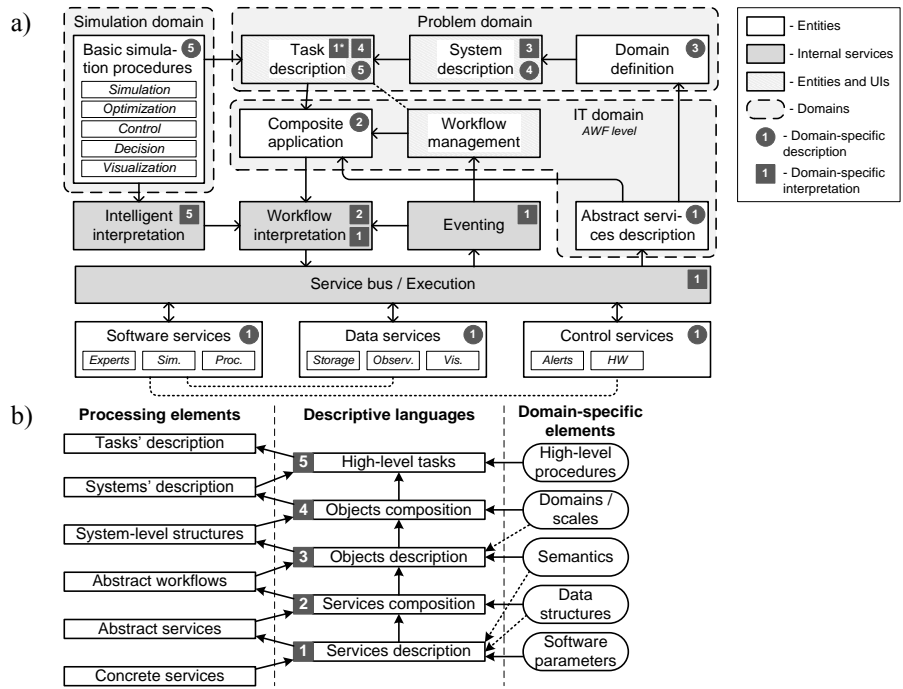

**Fig. 2** Computational environment a) high-level architecture; b) descriptive languages

5. High-level tasks description define additional techniques which can be applied to the system's structure, described earlier. Processing of the system's structure can be performed for the different purposes: visualization, parameters optimization, decision support etc. The simplest way of this technique implementation is development of heuristics or rule-based extension to the basic knowledge base.

## 4 Implementation Details

The proposed architecture was implemented on the basis of CLAVIRE cloud computing platform [9], which allows building composite applications using domain specific software available within the distributed environment. The platform implements the iPSE concept, thus, the knowledge based solution for composite application development and execution can be integrated into existing solution.

## *4.1 Basic Knowledge Structure*

The core conceptual and technological solution for continuous integration is hierarchy of concepts within simulation process:

*1 – Simulated object*, which represents the main entity being explored. The object can be considered as a composite entity, or system of objects.

*2 – Simulated model*, which describes a set of static and dynamic characteristics of the object and can be used to explore it.

*3 – Method* can be defined as an imperative description of the model usage process. Methods are implemented in different simulation software.

*4 – Software packages* are used as algorithmic implementations of the defined methods. Usually this software is developed by the domain specialists.

*5 – Service* within a distributed computational environment (in case we are using SOA) can be considered as the software deployed on computational resource.

This hierarchy is developed to integrate the domain-specific concepts (1-3) and technological concepts (4-5). It is the core concept for development of knowledge-based solutions, which support high-level task definition, which in turn can be automatically translated using interconnection between concepts of the hierarchy.

To implement the proposed conceptual hierarchy the knowledge base in a form of ontological structure was developed within a framework of Virtual Simulation Objects (VSO) technology [10]. This ontological structure implements domain-specific concept of the hierarchy (1-3) and their interconnection. The technological concepts (4-5) were described as links to corresponding knowledge expressed using a set of DSLs (see Section 4.2). The objects can be interconnected with a help of VSO technology to form the system's semantic description. The VSO technology presents instrumental environment with graphical interface (see Section 4.3) for building a system's description using the library of objects. Thus, the VSO presents an expressive technology for objects description; composition and high-level task definition (see fig. 2b). The expressive technology includes: a) descriptive languages with graphical notation for objects description and composition; b) tools to design a composite solution for complex system's simulation; c) interpretation engine for translation the descriptions into the AWF form.

## *4.2 Domain-Specific Languages*

Domain-specific languages (DSL) [11] present a powerful technology for building expressive tools within different problem domains. CLAVIRE environment uses a couple of related DSLs [9]:

**EasyPackage** – a language for services description based on Ruby language. It is used to define the knowledge on software services: input and output parameters, used data and files structures etc. A specific part of knowledge, expressed using this language is parametric performance models, used to estimate and to optimize performance characteristics of the services within cloud computing environment [12]. Also the language provides the user with ability to describe semantic meaning of defined entities within the problem domain. Considering the conceptual hierarchy presented in Section 4.1 this language describe the knowledge of level 4. The knowledge on last (5) level of this hierarchy is described by the means of ResourceBase system within CLAVIRE environment which includes the description of particular services (using JSON format) linked with the software description in the EasyPackage language and used within the knowledge base.

**EasyFlow** – a language for composite application description in a form of AWF. The WF is a central object within the CLAVIRE environment which defines the structure of composite application. It is developed using ANTLR framework and uses the software description presented using EasyPackage language to define software calls. So, the language presents expressive tool for com-posite application description. The interpretation of the AWF described by the lan-guage includes searching for appropriate services, processing of the incoming and outgoing data, tuning services' execution parameters. These procedures also can be performed using the knowledge presented using EasyPackage language.

*4.3 User Interfaces*

A knowledge-based approach can be used to develop user interfaces (UI) which support human-computer interaction with required level of abstraction. Several classes of UI were developed using CLAVIRE environment:

**UI 1** The high-level interfaces can be developed using VSO technology. Knowledge on the simulation of particular objects can be used to develop domain-specific graphical language to define simulated objects and systems. Fig. 3a presents the UI of VSO-based workspace which can be used to define simulated system as a composition of objects available within a library. The designed description can be translated into a WF and executed as a regular composite application.

**UI 2** The domain-specific support of composite application development can be performed using the interfaces which provide the user with qualitative comparison of available solutions. Fig. 3b shows a tree-based UI which uses the concepts of levels (2-4) within conceptual hierarchy to compare different solutions for particular domain tasks [13]. The user can define the parameters of the task and the system will compare available solutions using a set of qualitative criteria.

**UI 3** WFMS usually provide the user with an interface for graphical (typically in a form of graph) or textual description of WF. Most WFMSes have interfaces common to Integrated Development Environments (IDEs) including the capability to manage project files, run the WF, monitor the execution process, view the results, etc. CLAVIRE environment also uses graphical UI based on EasyFlow language (both with textual and graphical notation) to define the AWF structure.

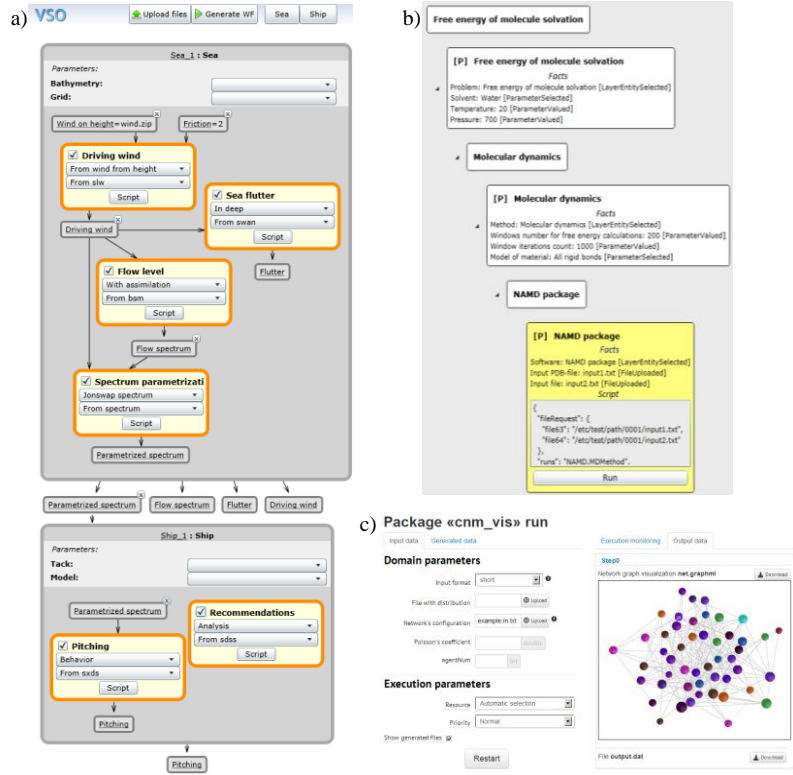

Fig. 3. Knowledge-based user interfaces

**UI 4** Problem-oriented interfaces (fig. 3c) allow the user to input all the required parameters for running particular task within cloud computing environment. The system performs automatic input checking; fulfill all the necessary procedures to translate input data into required format. Also the results of the execution can be obtained and viewed using the same interface. This UI is generated automatically using the EasyPackage description.

**UI 5** The lowest level of the UI is provided by a command-line interface and API library to access the computational environment. This type of interface is rarely used (mainly for the purpose of software integration).

### 4.4 Touch Table Application: an Example

Among the other tasks, which require simulation, decision support system can be mentioned as a complex example which involves different type of users interacting with the system and with each other. An application for collaborative decision support for the task of surge floods prevention in St. Petersburg [14] was developed using touch table hardware for collaboration (fig. 4a). While the main interface of this application presents geographic information system based on the map of sea near the St. Petersburg city (fig. 4b) the decision making process can be supported by the mean of additional expressive technologies mentioned earlier: a) WFMS (UI 3) is used to run simulation and decision making WFs, which provide the analysis of the forecast of sea level and possible plans to prevent the flood; b) problem-oriented interfaces (UI 4) is used for more detailed simulation using the available software; c) high-level knowledge-based interfaces (UI 1-2) are used to build, extend and analyze different scenarios of the situation development during the collaborative decision making; d) finally the API (UI 5) is used to integrate the parts of the solution (touch table, decision makers UIs) with CLAVIRE environment. The DSLs are used within the application to express the knowledge of experts (in form of predefined simulation WFs) and description of available cloud services, while ontological structure can be used to support semantic integration of different parts within the solution.

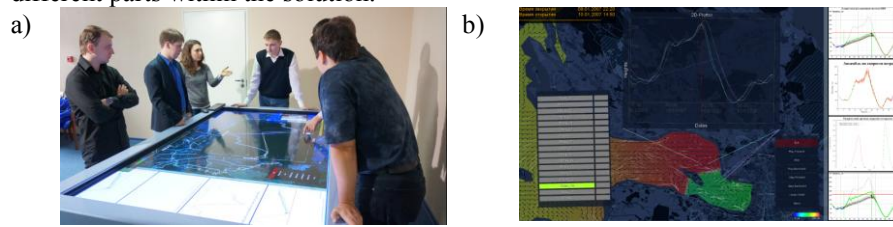

**Fig. 4.** Touch table application a) collaborative decision making; b) main user interface

## 5 Discussion and Conclusion

Solving contemporary e-Science tasks requires involvement of knowledge from different domains: a set of problem domains should be extended with IT domain (as the underlying software often is characterized by a complex usage), simulation domain (as a lot of techniques for complex system simulation require specific knowledge). All these knowledge needs to be formalized and dynamically integrated because the parts of knowledge might be updated separately. Today the ontology is often used as an integration technology (see, e.g. [15]). On the other hand, DSLs [11] present powerful technology for declarative and imperative knowledge expression using graphical of textual notation. As a result, the combination of these technologies can be used as a core technological solution for comprehensive knowledge-based support of human-computer interaction.

Presented work is devoted to development of the expressive toolbox which will support the processes of expression and interpretation in the most convenient and easiest way for the user. The continuous integration of different technologies using core ontological structure allows automatic interpretation and translation of the information provided by the user on different levels of abstraction. The developed expressive toolbox was implemented using CLAVIRE cloud computing environment.

**Acknowledgements** This work is supported by the projects "Technology of system-level design and development of inter-disciplinary applications within cloud computing environment" (agreement 14.B37.21.1870), "Virtual testbed for complex system supercomputing simulation" (agreement 14.B37.21.0596) granted from the Ministry of Education and Science of the Russian Federation and project "Urgent Distributed Computing for Time-Critical Emergency Decision Support" performed under Decree 220 of Government of the Russian Federation.